\documentclass[aps,prx,reprint,superscriptaddress,amsmath,amssymb,longbibliography]{revtex4-2}

\usepackage{graphicx}
\usepackage{xcolor}
\definecolor{magmacoral}{HTML}{F4675C}
\usepackage[colorlinks=true,linkcolor=magmacoral,
citecolor=magmacoral,urlcolor=magmacoral]{hyperref}

\begin{document}

\title{Anomaly-ranked confirmation slows new-class learning}

\author{Argyro Sasli}
\affiliation{School of Physics and Astronomy, University of Minnesota, Minneapolis, MN 55455, USA}
\author{Michael Coughlin}
\affiliation{School of Physics and Astronomy, University of Minnesota, Minneapolis, MN 55455, USA}

\date{\today}

\begin{abstract}
Anomaly detectors rank unfamiliar examples for expert confirmation, and the confirmed examples teach a classifier to recognise a new class. Across five benchmarks from physics, medicine and computer vision, the members flagged by the detector are those the classifier learns first. In three systems, confirming the highest-ranked members of a new class gives lower recognition than confirming the same number of random members. Handing subsequent selection to the promoted classifier recovers part of this loss in two systems. A paired intervention further shows that replacing nearby known classes with distant ones sharply reduces the confirmations needed for recognition. These results distinguish finding a new class from selecting examples that teach its extent: within the tested classes, concentrating confirmations at the top of the anomaly ranking makes learning less efficient.
\end{abstract}

\maketitle

\section*{Introduction}

Anomaly detection is a route to discovering unfamiliar events. In time-domain astronomy, gravitational-wave detector characterisation and clinical monitoring, discovery pipelines pair a classifier trained on known classes with an anomaly detector that flags unfamiliar inputs \cite{malanchev2021ztf,lochner2021astronomaly,ishida2021active,soni2021gravityspy,sasli2026astranet,companion}. Experts label the selected candidates through spectroscopy, visual inspection or clinical review, and we call each such label a confirmation. Confirmed examples enter training as a new class, but telescope time and expert hours limit how many can be obtained.

Out-of-distribution detection and open-set recognition evaluate whether a system flags unfamiliar inputs \cite{scheirer2013openset,bendale2016openmax,hendrycks2017baseline,liu2020energy,yang2022openood,vaze2022openset}. Few-shot, class-incremental and novel-category learning evaluate how a system learns from labelled examples \cite{wang2020fewshot,rebuffi2017icarl,han2019ncd,vaze2022gcd,viering2023curves}. Active learning addresses the value and selection bias of queried examples \cite{settles2009active,dasgupta2008hierarchical}, including pools containing unknown classes \cite{ning2022openset,schmidt2025joda}. Here we examine the link between discovery and subsequent learning: does the ranking that finds a new class select useful examples for teaching that same class?

Our FLARE transient pipeline \cite{companion} measures recognition after returning confirmed examples of a withheld class to training. Those examples were drawn independently of the anomaly ranking. In an anomaly-driven follow-up programme, however, the detector also influences which members receive labels.

We test that choice by withholding classes from five labelled benchmarks and returning their members as confirmations, using benchmark labels in place of expert labels. We first compare which members are flagged and learned, then intervene on confirmation selection and on the known classes surrounding the new class.

\begin{figure*}[t]
\centering
\includegraphics[width=0.92\textwidth]{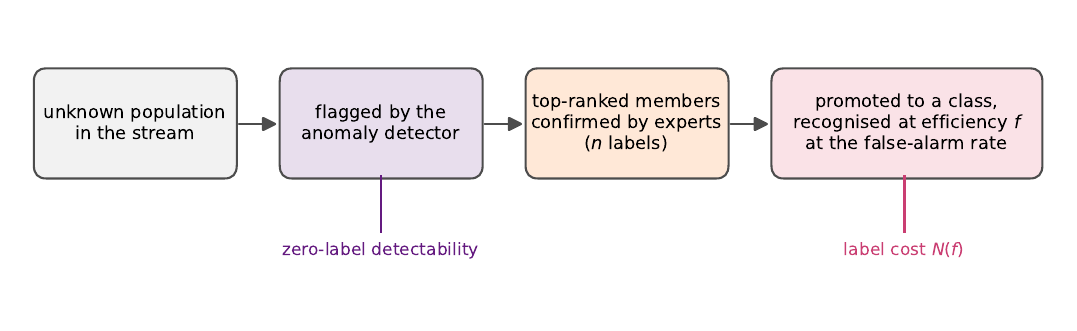}

\medskip
\includegraphics[width=\textwidth]{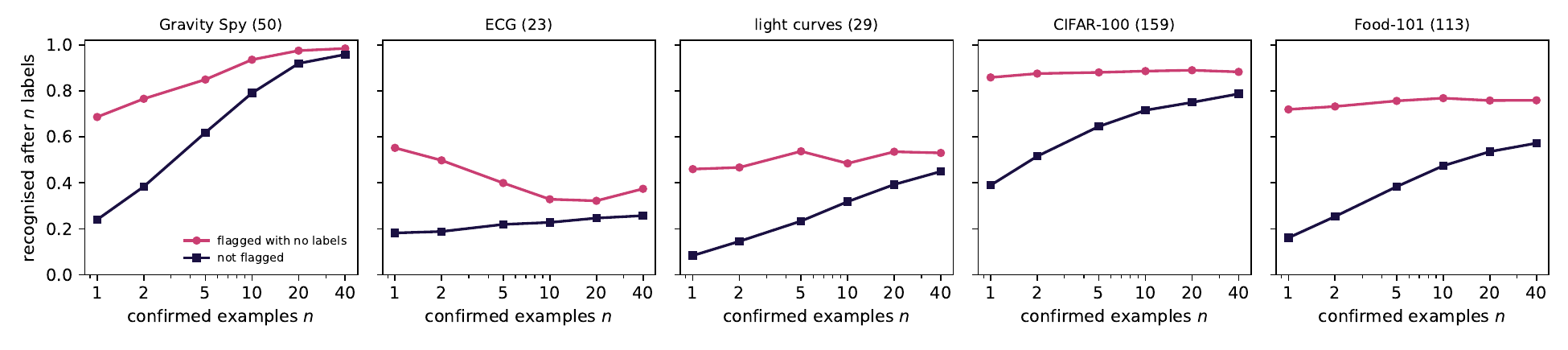}
\caption{
\textbf{Discovery and learning favour the same members of a new class.}
\emph{Top:} an anomaly detector flags an unfamiliar population, experts confirm selected members, and the confirmed examples enter training as a new class. Recognition is measured at a fixed false-alarm rate; the label cost $N(f)$ is the estimated number of confirmations needed to reach efficiency $f$.
\emph{Bottom:} probability that a held-out member is recognised at a $1\%$ false-alarm rate after promotion with $n$ random confirmations, separately for members flagged before any label (magenta) and members not flagged (dark). Curves average over withholds with at least three members in each group; parentheses give the number of withholds.
}
\label{fig:lifecycle}
\label{fig:member}
\end{figure*}

\section*{Results}

\subsection*{The detector flags the members the classifier learns first}

For each eligible withheld class, we divided its held-out members into those flagged by the zero-label anomaly detector and those not flagged. Both groups were evaluated with the same classifier, promoted using random confirmations. Differences therefore concern which members are recognised, rather than differences in the confirmations used for training.

Across all five datasets, flagged members are recognised after fewer confirmations than unflagged members (Fig.~\ref{fig:member}). In Gravity Spy, recognition after one confirmation is $0.69$ for flagged members and $0.24$ for unflagged members; the groups converge only near forty confirmations. In CIFAR-100 and Food-101, flagged members reach $0.86$ and $0.72$ after one confirmation, whereas unflagged members reach $0.79$ and $0.57$ after forty. In the electrocardiograms, recognition of flagged members declines as confirmations accumulate but remains above recognition of unflagged members.

The detector thus selects a region of the new class that the classifier can already learn from very few confirmations. This motivates testing whether repeatedly confirming that region is an efficient way to learn the rest of the class.

\subsection*{Detector-selected confirmations slow new-class learning}

For each withheld class, we compared three choices of $n$ confirmations: its highest anomaly scores, $n$ random members, and its lowest anomaly scores. Only confirmation selection changed; the learner and evaluation protocol were fixed. We tested Gravity Spy with gradient-boosted trees on image features, CIFAR-100 with a linear classifier on ResNet-50 features, and electrocardiograms with gradient-boosted trees on hand-crafted features.

\begin{figure*}[t]
\includegraphics[width=\textwidth]{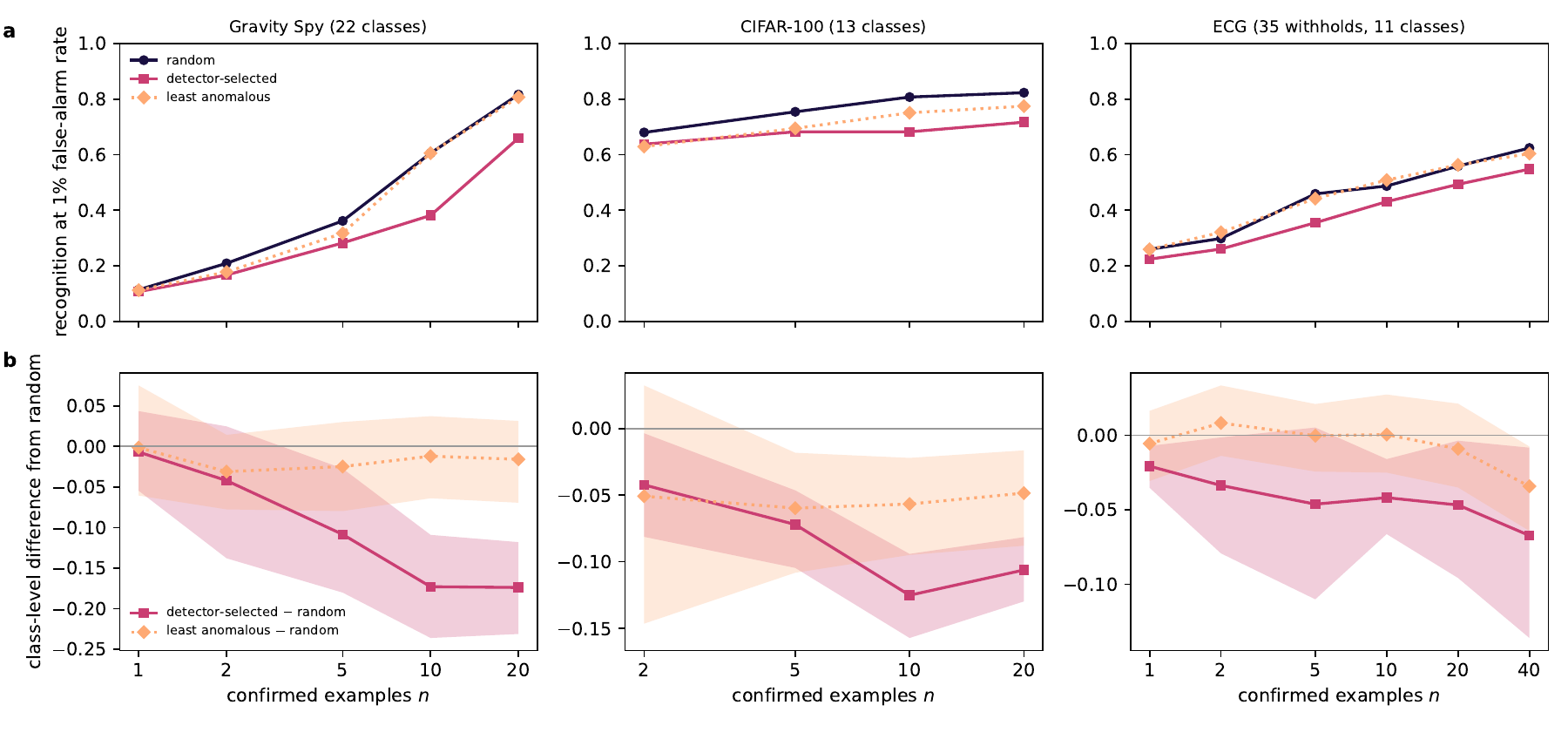}
\caption{
\textbf{Detector-selected confirmations slow new-class learning.}
\emph{(a)} Mean recognition at a $1\%$ false-alarm rate after promotion with $n$ random members (dark), the $n$ highest-ranked members under the anomaly detector (magenta), or the $n$ least anomalous members (light). Gravity Spy uses all twenty-two classes; CIFAR-100 uses thirteen classes with zero-label detectability between $0.1$ and $0.9$; ECG uses thirty-five withholds of eleven classes in geometry-selected sub-taxonomies.
\emph{(b)} Class-averaged differences from random selection with $95\%$ bootstrap intervals over classes. Gravity Spy at five, ten and twenty confirmations averages three training seeds. Each ECG class enters once, through its median over the sub-taxonomies containing it.
}
\label{fig:selection}
\end{figure*}

Detector-selected confirmations give lower recognition than random confirmations in all three systems (Fig.~\ref{fig:selection}). After ten confirmations, the difference is negative for twenty of twenty-two Gravity Spy classes, all thirteen CIFAR-100 classes and ten of eleven ECG classes. Class-bootstrap intervals exclude zero from five confirmations onwards in Gravity Spy, at every measured budget in CIFAR-100 and at five of six budgets in the ECGs. Averaged over three Gravity Spy training seeds, the difference at ten confirmations is $-0.17$ ($95\%$ interval $-0.24$ to $-0.11$). At twenty confirmations, recognition remains $0.66$ against $0.82$ for random selection.

The penalty is not confined to a few extreme examples. Drawing $n$ confirmations at random from the $2n$ highest-ranked members also underperforms unrestricted random selection in all eight Gravity Spy classes tested. Selecting the least anomalous members gives recognition close to or slightly below random selection. These comparisons are consistent with the detector concentrating confirmations on members the classifier already recognises readily, while a random sample covers a broader part of the class.

To match the recognition obtained from ten random confirmations, detector-selected confirmations require approximately eighteen in Gravity Spy and nineteen in the ECGs, estimated by interpolation of the mean curves. In CIFAR-100, twenty detector-selected confirmations remain below the recognition obtained from five random ones. Random-confirmation benchmarks can therefore underestimate the label requirement under anomaly-ranked confirmation in these systems.

\begin{figure*}[t]
\includegraphics[width=\textwidth]{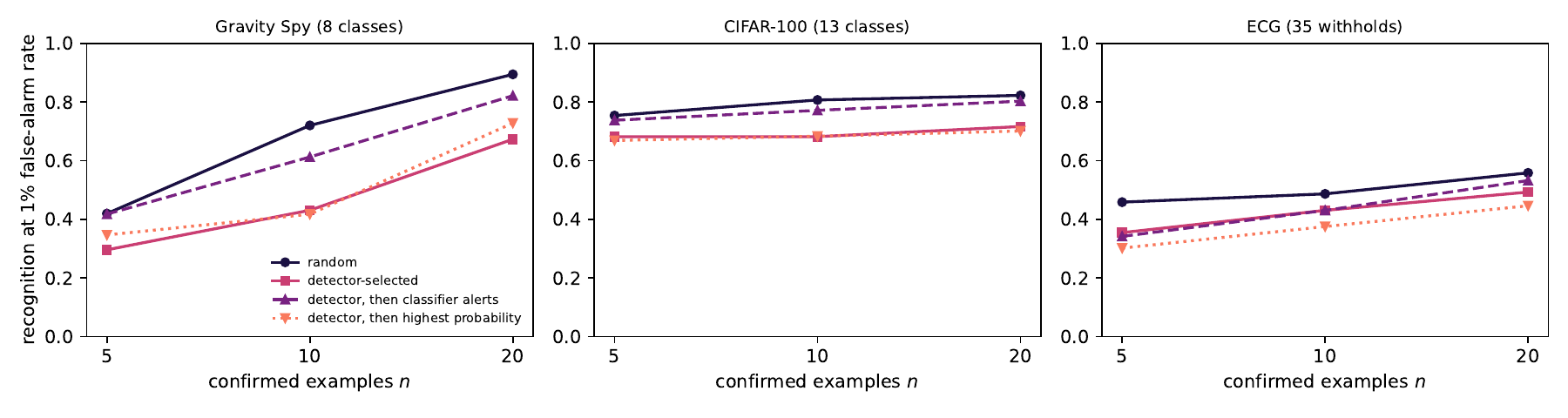}
\caption{
\textbf{Handing later confirmations to the classifier recovers part of the penalty.}
Recognition at a $1\%$ false-alarm rate for the hand-off cohorts: eight Gravity Spy classes with zero-label detectability between $0.1$ and $0.9$, thirteen CIFAR-100 classes, and thirty-five ECG withholds involving eleven classes. Following two detector-selected confirmations, the promoted classifier selects further members by sampling among its alerts (dashed) or by highest promoted-class probability (dotted). Random and detector-selected policies are evaluated on the same cohorts.
}
\label{fig:handoff}
\end{figure*}

A classifier can recover part of the loss by taking over subsequent selection (Fig.~\ref{fig:handoff}). We first confirmed the detector's two highest-ranked members, promoted the class, and then sampled further confirmations among members flagged by the promoted classifier. At ten confirmations, this recovers $63\%$ of the gap between detector-selected and random selection in Gravity Spy and $72\%$ in CIFAR-100. It gives no corresponding improvement in the ECGs. Choosing only the members with the highest promoted-class probability performs worse than sampling among the classifier's alerts (Supplementary Sec.~\ref{si:handoff}).

The same ordering holds when confirmations are drawn from a mixed stream that also contains a second withheld class and known-class events. At twenty confirmations, the detector's highest-ranked events include more members of the new class than random draws among its alerts, $15.6$ against $12.9$ in Gravity Spy and $10.8$ against $7.2$ in CIFAR-100, yet give lower mean recognition, $0.57$ against $0.66$ and $0.70$ against $0.72$. Handing selection to the promoted classifier gives the highest recognition in both systems, $0.68$ and $0.77$ (Supplementary Sec.~\ref{si:mixed}).

\subsection*{Known-class context changes the cost of learning}

We tested the influence of neighbouring known classes with a paired intervention. For each of twenty-one Gravity Spy classes, two six-class taxonomies shared the withheld class, two background classes and the confirmed examples. The remaining three classes were either its nearest or its farthest eligible neighbours in feature space.

\begin{figure*}[t]
\includegraphics[width=\textwidth]{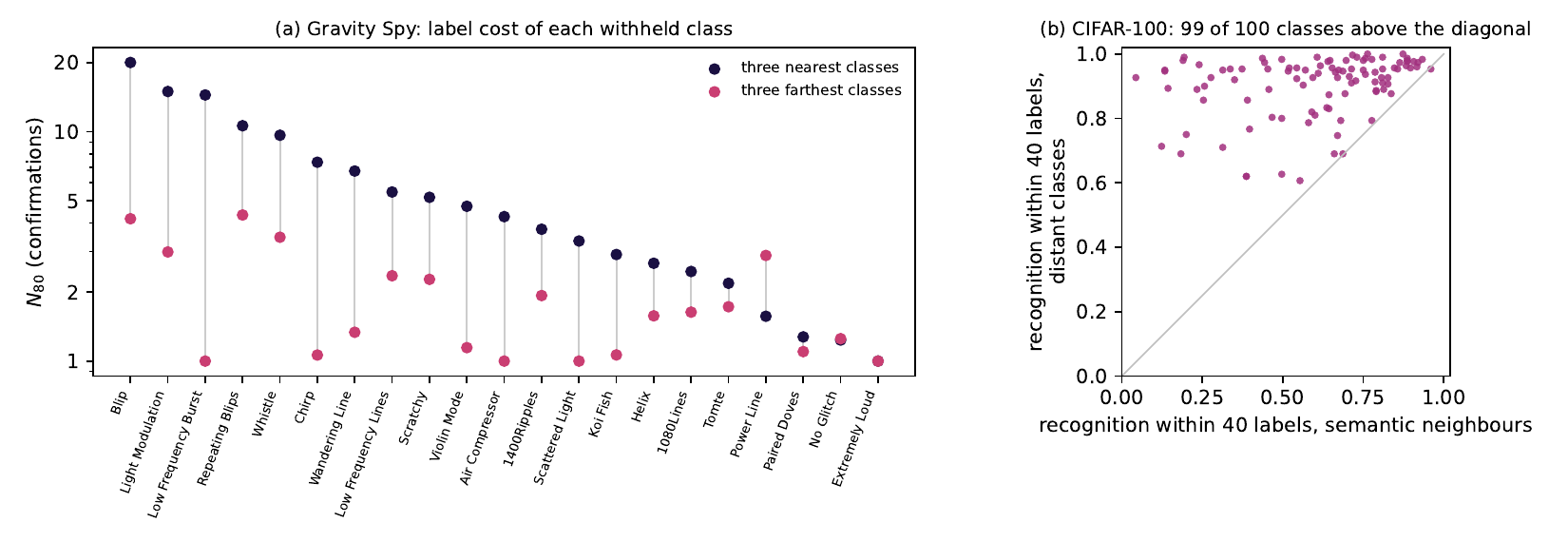}
\caption{
\textbf{Known-class context changes new-class learning cost.}
\emph{(a)} Estimated confirmations needed for $80\%$ recognition at a $1\%$ false-alarm rate for each Gravity Spy class, among its three nearest eligible known classes (dark) or its three farthest (magenta). Two further background classes are shared. Costs are computed separately in four replicates and averaged within each class.
\emph{(b)} CIFAR-100: best recognition through forty confirmations with semantic neighbours versus distant classes. Five further background classes and the confirmation draws are shared; ninety-nine of one hundred points lie above the diagonal.
}
\label{fig:swap}
\end{figure*}

Distant neighbours reduce the estimated number of confirmations needed for $80\%$ recognition for eighteen of twenty-one classes. The median falls from $4.3$ to $1.6$ confirmations (Fig.~\ref{fig:swap}a). The result persists across four replicate draws and alternative readings of label cost (Supplementary Sec.~\ref{si:swap}).

In CIFAR-100, we replaced a withheld class's four semantic siblings with four classes from other superclasses, holding five further background classes and the confirmation draws fixed. Best recognition through forty confirmations is higher with distant classes for ninety-nine of one hundred classes (Fig.~\ref{fig:swap}b). These interventions establish a dependence on known-class context; they do not establish a geometric law predicting the size of the reduction.

\section*{Discussion}

The examples that reveal a new class need not be those that teach its extent. In the tested systems, anomaly-ranked confirmation concentrates labels on members the promoted classifier recognises readily, producing lower recognition than random confirmation at the same budget. Handing later selection to the classifier's alerts recovers part of this loss in two systems, while selecting only its highest-probability members remains ineffective.

These results distinguish novelty from representativeness within a new class. Programmes such as the Bright Transient Survey select confirmations through a predefined brightness-based selection function rather than anomaly rank \cite{fremling2020bts,perley2020bts}. Combining anomaly-driven discovery with broader confirmation sampling may similarly help pipelines learn newly identified populations.

The main experiments draw confirmations from members of the withheld class alone. In a mixed stream, where labels are also spent on known classes and a second unknown class, the detector finds more members of the new class and still teaches it less efficiently than random sampling among its alerts. Withholding an existing class also differs from encountering a genuinely new population. Noisy labels and unequal confirmation costs may alter the trade-off.

The neighbour intervention shows that label requirements depend on the surrounding taxonomy as well as on the class itself. It does not supply a general rule for predicting how many labels a class needs. A prospective evaluation should record candidate rankings, all confirmation costs and subsequent recognition, testing whether broader sampling retains its advantage under expert confirmation.

\section*{Methods}

\subsection*{Datasets and representations}

We used five labelled benchmarks with fixed feature representations and train, validation and test partitions.

\emph{Gravity Spy} contains twenty-two morphological classes of LIGO detector glitches \cite{zevin2017gravityspy}. Models use $984$ pooled-pixel features from four spectrogram windows and a stratified $70/15/15$ split. Before withholding, the pipeline passed its specified accuracy criterion of $0.90$, reaching $0.963$. We removed all $37$ records belonging to $18$ exact-duplicate groups and repeated the experiments with otherwise unchanged designs and seeds. Four groups had copies across the test and training or validation partitions. All reported Gravity Spy results use the deduplicated data.

\emph{PTB-XL} provides twelve-lead electrocardiograms \cite{wagner2020ptbxl}. We retained records with a unique dominant diagnostic subclass at annotation likelihood $\geq 50$, giving $15\,041$ of $21\,799$ records. Models use $108$ per-lead summary features: moments, amplitude range, autocorrelation-based rhythm measures and spectral-band fractions. Official folds $1$--$8$ supply training members, fold $9$ validation and thresholds, and fold $10$ evaluation. Over sixteen subclasses with at least fifty members, full-taxonomy macro AUC was $0.855$, above the specified $0.80$ criterion. Restricting to unambiguous labels limits extrapolation to mixed clinical cases.

\emph{Optical transients} use $160$ class-agnostic light-curve features for $10\,338$ quality-passing Bright Transient Survey objects and $37$ classical novae, with the splits documented by the companion FLARE study \cite{companion,fremling2020bts,perley2020bts}. Host and redshift context are excluded. Full-taxonomy macro AUC over sixteen classes was $0.909$, above the specified $0.80$ criterion.

\emph{CIFAR-100} has one hundred classes grouped into twenty superclasses \cite{krizhevsky2009cifar}. We used ImageNet-pretrained ResNet-50 features \cite{he2016resnet}, reduced from $2048$ to $256$ principal components fitted without labels on training images. Each class contributes $450$ training, $50$ validation and $100$ test images. Full-taxonomy macro AUC was $0.990$, above the specified $0.80$ criterion.

\emph{Food-101} has one hundred and one classes \cite{bossard2014food}. We sampled $300$ training and $50$ validation images per class from its training split and $100$ test images per class from its official validation split, with seed $0$. Images were resized to $256$ pixels on the shorter side and centre-cropped to $224$ before feature extraction.

The member-level comparison includes full and sparse taxonomies. Gravity Spy uses four random six-class and three eleven-class taxonomies; ECG and light curves use four six-class and three nine-class taxonomies, retaining the normal ECG or Type Ia supernova background class. CIFAR-100 and Food-101 use six ten-class and three twenty-class taxonomies. Random taxonomy draws use seed $11$.

\subsection*{Models and withholding}

Gravity Spy, ECG and transient models use multiclass LightGBM \cite{ke2017lightgbm} with $63$ leaves, learning rate $0.08$ for Gravity Spy and $0.06$ elsewhere, and early stopping after $80$ validation rounds. Maximum rounds are $1500$ and $1200$, respectively. Gravity Spy uses feature and bagging fractions $0.7$ and $0.8$. Image models use multinomial logistic regression on frozen features, fitted by full-batch L-BFGS with an $L_2$ penalty of $10^{-4}$. Except in Gravity Spy, class weights are proportional to $\sqrt{N/(K N_k)}$, where $N_k$ is the number of training members of class $k$ among $K$ classes.

For each withheld class $c$, we trained on the other $K-1$ classes. The zero-label anomaly score is the energy of the known-class logits \cite{liu2020energy},
\begin{equation}
E(x)=-\log\sum_k e^{z_k(x)}.
\end{equation}
Its threshold is the $99$th percentile of scores on held-out known-class events. Zero-label detectability is the fraction of held-out members of $c$ above this threshold.

We then added $n$ confirmed members of $c$ as a new training class and retrained. Recognition efficiency is the fraction of held-out target members whose promoted-class probability exceeds its $99$th-percentile threshold on held-out known-class events. Thresholds are estimated separately for each promoted model. The nominal $1\%$ false-alarm rate is a benchmark operating point, not a guarantee for a future stream. Target evaluation members are excluded from confirmation selection and training.

Random-confirmation curves average three draws per label count. The member-level analysis uses raw recognition at each measured budget. For the Gravity Spy swap, $N_{80}$ is the first crossing of $80\%$ on the running maximum of the mean recognition curve, linearly interpolated between measured budgets. If the first measured budget reaches the target, its label count is used; classes not reaching it are censored. Fractional costs are estimated crossings, not guaranteed integer requirements. Alternative readings are reported in Supplementary Sec.~\ref{si:swap}.

\subsection*{Member-level comparison}

For each withhold, we used the zero-label detector to partition held-out target members into flagged and unflagged groups. Eligible withholds contain at least ten held-out members and at least three in each group. Recognition after random confirmation is averaged over three training draws and then over withholds. The comparison includes $50$ Gravity Spy, $23$ ECG, $29$ light-curve, $159$ CIFAR-100 and $113$ Food-101 withholds. This exploratory analysis preceded and motivated the confirmation-policy experiments.

\subsection*{Confirmation selection and hand-off}

Candidates are members of the withheld class available for promotion: training and validation members in Gravity Spy, training images in CIFAR-100, and folds $1$--$8$ in PTB-XL. Policies select the highest-energy candidates, the lowest-energy candidates, or random candidates. A stochastic detector policy samples $n$ members from the $2n$ highest-energy candidates. Random policies average three draws.

The selection experiment uses all twenty-two Gravity Spy classes through twenty confirmations; thirteen CIFAR-100 classes with stored zero-label detectability between $0.1$ and $0.9$ through twenty confirmations; and thirty-five ECG withholds involving eleven classes through forty confirmations. CIFAR-100 eligibility was fixed before the selection runs. ECG uses seven geometry-selected six-class taxonomies because no class in its full taxonomy has zero-label detectability above $0.1$. Their construction is described in Supplementary Sec.~\ref{si:selection}.

Hand-off policies first confirm the detector's two highest-ranked members. At budgets $2$, $5$ and $10$, the refitted classifier selects the next confirmations from the remaining pool by sampling among its alerts, by highest promoted-class probability, or by probability closest to its threshold. These experiments use eight Gravity Spy classes with detectability between $0.1$ and $0.9$, the thirteen CIFAR-100 classes and the thirty-five ECG withholds.

The mixed-stream test withholds a second class, drawn at random among classes with at least fifty candidates, and moves $20\%$ of each known class's training members into the stream. Confirmations of the second class or of known classes consume budget without entering training. The detector confirms the highest-energy stream events; the alerts policy samples among stream events above the $1\%$ energy threshold; the hand-off policy follows the detector until at least two target members are confirmed and then samples among the promoted classifier's alerts. Budgets are $5$, $10$, $20$ and $40$ confirmations, with three draws per randomised policy, for the eight Gravity Spy hand-off classes and the thirteen CIFAR-100 classes.

Equivalent confirmation counts are interpolated in $\log n$ from the running maximum of the mean detector-selected curve. They compare equal recognition levels and are approximate.

\subsection*{Neighbour intervention}

Gravity Spy neighbours are ranked on training data by directional isolation (Supplementary Sec.~\ref{si:swap}). Eligible background classes have at least $100$ training and $40$ calibration members. For each target, the two six-class taxonomies contain the target, two shared classes at the median rank, and either its three nearest or three farthest remaining candidates. Background classes are subsampled to equal training and calibration counts; shared subsamples and confirmed-member indices are identical across arms.

The comparison uses four replicate draws. We compute $N_{80}$ separately in each replicate and average within each class before taking the median across classes. Protocol thresholds are recomputed in each arm; a diagnostic comparison uses identical calibration members from the shared classes.

In CIFAR-100, the near arm contains the target's four semantic siblings, the far arm four classes from different superclasses, and both share five further background classes ($K=10$). Training counts and confirmation draws are paired. The reported endpoint is best recognition through forty confirmations.

\subsection*{Statistics and reproducibility}

Confirmation-policy differences are summarised at class level. Each ECG class contributes its median over the sub-taxonomies containing it. Reported $95\%$ intervals are percentile bootstrap intervals over classes; they do not include uncertainty from threshold estimation or new confirmation draws. Gravity Spy results at five, ten and twenty confirmations average training seeds $0$, $1$ and $2$. Seed sensitivity and failed criteria are reported in Supplementary Sec.~\ref{si:selection}.

Neighbour-swap sign tests are paired, one-sided and censoring-aware, excluding ties. Spearman correlation describes the relation between isolation changes and cost changes. No multiple-comparison correction was applied. Sample sizes follow benchmark eligibility rather than a power calculation.

The member-level analysis was exploratory. Predictions for subsequent analyses were recorded in dated scripts before their corresponding measurements, without an external registry. Earlier undated file timestamps do not establish prior specification. Declaring scripts, digests and the full ledger, including failed predictions, accompany the code.

Runs used Python $3.12$, LightGBM $4.6.0$, scikit-learn $1.8.0$, NumPy $2.3.5$, SciPy $1.17.1$, PyTorch $2.10.0$ and torchvision $0.25.0$. Feature extraction used a GPU and subsequent experiments used CPU nodes on NCSA Delta. Exact seeds, preprocessing and run configurations are recorded in the scripts.

\section*{Data availability}

The transient benchmark is released with the companion paper \cite{companion} at \url{https://github.com/applecider-ml/flare}. Gravity Spy \cite{zevin2017gravityspy}, CIFAR-100 \cite{krizhevsky2009cifar}, Food-101 \cite{bossard2014food} and PTB-XL \cite{wagner2020ptbxl} are public; PTB-XL is available at \url{https://physionet.org/content/ptb-xl/}. Derived features, embeddings and per-class outputs accompany the code.

\section*{Code availability}

Feature extraction, withholding and promotion, confirmation policies, controls, prediction records and figure-generation scripts are available at \url{https://github.com/asasli/label-cost}.

\begin{acknowledgments}
The authors acknowledge support from the National Science Foundation through grants PHY-2117997, PHY-2308862 and PHY-2409481. This work uses observations from the Zwicky Transient Facility and its Bright Transient Survey; Gravity Spy classifications contributed by Zooniverse volunteers and LIGO detector-characterisation data; and PTB-XL data made available through PhysioNet by the Physikalisch-Technische Bundesanstalt.

During manuscript preparation, A.S. used Claude (Opus 5.5) to improve English language, clarity, and organization. The authors reviewed and edited all AI-assisted text and take full responsibility for the final manuscript.
\end{acknowledgments}

\bibliography{references}

\clearpage
\onecolumngrid
\begin{center}
\Large\textbf{Supplementary Information}
\end{center}
\twocolumngrid

\setcounter{table}{0}
\setcounter{figure}{0}
\setcounter{section}{0}
\renewcommand{\tablename}{Supplementary Table}
\renewcommand{\figurename}{Supplementary Figure}
\renewcommand{\thetable}{\arabic{table}}
\renewcommand{\thefigure}{\arabic{figure}}
\renewcommand{\thesection}{S\arabic{section}}

\section{Confirmation-policy robustness}
\label{si:selection}

\subsection{ECG taxonomy selection}

ECG sub-taxonomies were constructed without using recognition outcomes. Starting from each pathological seed class, we greedily added the class maximising the minimum pairwise directional isolation until five pathological subclasses joined the normal class. We retained the seven distinct sets with the highest minimum isolation and withheld each pathological member in turn. The resulting thirty-five withholds involve eleven distinct classes. Class-level summaries keep these repeated appearances together.

This design conditions the ECG result on geometrically separated taxonomies. The originally planned intermediate-detectability comparison could not be evaluated in the full ECG taxonomy because no class had detectability above $0.1$.

\subsection{Training seeds}

Gravity Spy's feature and row subsampling make its learner sensitive to the training seed. With seeds $1$ and $2$, detector-selected confirmation gives lower recognition than random confirmation after ten labels for seventeen of twenty-two classes in each case. Mean recognitions are $0.46$ against $0.57$ and $0.44$ against $0.62$, respectively.

Averaging seeds $0$, $1$ and $2$, the difference is $-0.17$ ($95\%$ class-bootstrap interval $-0.24$ to $-0.11$), negative for twenty of twenty-two classes. The predeclared criterion of at least eighteen negative classes in each additional seed was not met. The detector arm's across-seed standard deviation for a single class is $0.15$, exceeding the proposed stability criterion of half the mean gap ($0.09$). The average penalty is therefore more stable than individual-class estimates.

The ECG trees use neither feature nor row subsampling and give identical predictions across seeds. The CIFAR-100 linear head solves a convex objective from a fixed start.

\subsection{Selection controls}

Sampling $n$ confirmations at random among the $2n$ highest-ranked candidates retains the penalty relative to unrestricted random sampling in all eight Gravity Spy classes tested. This argues against the effect being confined to the most extreme candidates.

The least anomalous policy did not consistently outperform random selection. The prediction that the detector penalty would shrink from five to twenty confirmations also failed in Gravity Spy, where it grew. Persistence over that interval replicated in CIFAR-100 but not in the ECG sub-taxonomies. These results limit any universal claim about how the penalty changes with budget.

\section{Neighbour-swap controls}
\label{si:swap}

\subsection{Neighbour ranking}

Directional isolation is computed from standardised training features reduced to fifty principal components. For classes $c$ and $j$, with means $\mu_c,\mu_j$ and Ledoit--Wolf covariance estimates $S_c,S_j$, define
\begin{equation}
S_{\mathrm{pooled}}=\tfrac12(S_c+S_j),
\qquad
w=S_{\mathrm{pooled}}^{-1}(\mu_c-\mu_j).
\end{equation}
The ranking uses
\begin{equation}
\hat I_{\mathrm{dir}}(c,j)=
\frac{|w^\top(\mu_c-\mu_j)|}
{\sqrt{w^\top S_cw+w^\top S_jw}}.
\end{equation}
Training members of the withheld class are used to define this controlled intervention. It is an explanatory benchmark, not a geometry estimate available before confirming that class.

\subsection{Replicates, thresholds and cost definition}

With the paper's interpolated cost, distant neighbours are cheaper for eighteen classes and dearer for two, with one tie. The direction persists in each of the four replicates. Excluding the two classes most frequently chosen as distant neighbours also retains the effect.

\begin{table}[t]
\caption{
Gravity Spy neighbour-swap controls on deduplicated data. Counts exclude ties. Costs are computed per replicate and averaged within class.
}
\label{tab:swapcontrols}
\begin{ruledtabular}
\begin{tabular}{lc}
Comparison & Result \\
\colrule
Interpolated $N_{80}$, cheaper : dearer & $18:2$ \\
Median $N_{80}$, near $\rightarrow$ far & $4.3\rightarrow1.6$ \\
Replicate cheaper : dearer counts &
$15:5,\ 19:1,\ 17:1,\ 17:3$ \\
First measured crossing, cheaper : dearer & $19:1$ \\
Median first crossing, near $\rightarrow$ far & $6.2\rightarrow2.2$ \\
Two frequent far classes excluded & $20:0$ \\
\end{tabular}
\end{ruledtabular}
\end{table}

\begin{table*}[t]
\caption{
Mean recognition at a $1\%$ false-alarm rate for the hand-off cohorts. ``Least'' selects the lowest anomaly scores; ``alerts'', ``top'' and ``margin'' follow the two detector-selected confirmations. Randomised policies average three draws. The thirty-five ECG withholds involve eleven classes and are not independent class replicates.
}
\label{tab:handoff}
\begin{ruledtabular}
\begin{tabular}{lccccccc}
System & $n$ & Random & Detector & Least & Alerts & Top & Margin \\
\colrule
Gravity Spy (8 classes) & 5  & 0.42 & 0.30 & 0.48 & 0.42 & 0.35 & 0.42 \\
                       & 10 & 0.72 & 0.43 & 0.66 & 0.61 & 0.42 & 0.53 \\
                       & 20 & 0.89 & 0.67 & 0.79 & 0.82 & 0.73 & 0.83 \\
CIFAR-100 (13 classes)  & 5  & 0.75 & 0.68 & 0.69 & 0.74 & 0.67 & 0.75 \\
                       & 10 & 0.81 & 0.68 & 0.75 & 0.77 & 0.68 & 0.80 \\
                       & 20 & 0.82 & 0.72 & 0.77 & 0.80 & 0.70 & 0.83 \\
ECG (35 withholds)      & 5  & 0.46 & 0.35 & 0.44 & 0.34 & 0.30 & 0.37 \\
                       & 10 & 0.49 & 0.43 & 0.51 & 0.43 & 0.38 & 0.49 \\
                       & 20 & 0.56 & 0.49 & 0.56 & 0.53 & 0.45 & 0.57 \\
\end{tabular}
\end{ruledtabular}
\end{table*}

Using the first measured budget reaching $80\%$, rather than interpolation, gives the same direction with or without the running maximum. Computing cost from each class's mean curve instead of averaging replicate costs gives medians $4.19$ and $1.33$, with lower cost for seventeen of twenty-one classes.

With thresholds estimated from identical shared-class calibration members, recognition after five confirmations is higher among distant neighbours for sixteen classes and lower for three. The effect therefore remains under a common calibration reference.

The increase in directional isolation does not predict the size of the cost reduction across classes ($\rho=-0.04$). Predictions of a corresponding, consistently weaker response in the anomaly channel were not all met. Energy detectability rises for twelve of twenty-one classes on replicate means. The intervention supports an effect of taxonomy composition on learning, without establishing a scalar geometric predictor or an absence of change in detectability.

\section{Hand-off policies and their limits}
\label{si:handoff}

The first two confirmations are detector-selected. Subsequent choices sample among the promoted classifier's alerts, take its highest-probability members, or take members closest to its threshold. All policies draw only from the withheld class's candidate pool.

Table~\ref{tab:handoff} reports the results of this investigation. At ten confirmations, sampling among classifier alerts beats detector-only selection in seven of eight Gravity Spy classes, all thirteen CIFAR-100 classes and eighteen of thirty-five ECG withholds. The criterion of improvement in at least two thirds of the withholds was not met in the ECGs.

At twenty confirmations, highest-probability selection performs worse than sampling among alerts in five of eight Gravity Spy classes, thirteen of thirteen CIFAR-100 classes and twenty-seven of thirty-five ECG withholds. Margin selection approaches the random-policy mean in CIFAR-100 and ECG, but the corresponding predeclared random-level criterion passed only in ECG. These comparisons do not establish equivalence between policies or a universally effective hand-off rule.

\section{Mixed candidate streams}
\label{si:mixed}

The stream for each target class contains its candidate members, the candidate members of a second withheld class and $20\%$ of the training members of every known class. Only confirmed target members enter training and every confirmation counts against the budget.

\begin{table}[h]
\caption{
Mixed-stream confirmation. Mean recognition of the target class at a $1\%$ false-alarm rate, with the mean number of target members among the confirmations in parentheses. Randomised policies average three draws.
}
\label{tab:mixed}
\begin{ruledtabular}
\begin{tabular}{lcccc}
System & $B$ & Detector & Alerts & Hand-off \\
\colrule
Gravity Spy  & 5  & 0.23 (3.8)  & 0.31 (3.1)  & 0.23 (3.8) \\
(8 classes)  & 10 & 0.42 (7.8)  & 0.42 (6.2)  & 0.49 (7.0) \\
             & 20 & 0.57 (15.6) & 0.66 (12.9) & 0.68 (14.4) \\
             & 40 & 0.74 (29.1) & 0.79 (25.2) & 0.88 (31.3) \\
CIFAR-100    & 5  & 0.54 (3.2)  & 0.54 (1.7)  & 0.54 (3.2) \\
(13 classes) & 10 & 0.67 (5.9)  & 0.67 (3.4)  & 0.72 (6.5) \\
             & 20 & 0.70 (10.8) & 0.72 (7.2)  & 0.77 (13.7) \\
             & 40 & 0.72 (20.1) & 0.74 (14.5) & 0.83 (28.9) \\
\end{tabular}
\end{ruledtabular}
\end{table}

At twenty confirmations, the detector confirms more target members than the alerts policy for six of eight Gravity Spy classes and eleven of thirteen CIFAR-100 classes (Table~\ref{tab:mixed}). The alerts policy gives recognition at least as high as the detector for eight of eight and eight of thirteen classes, and hand-off exceeds the detector for seven of eight and eleven of thirteen. All four predeclared criteria were met. In CIFAR-100 the difference between the detector and alerts policies is small, and most of the gain comes from hand-off.

\end{document}